\documentclass[a4paper,twocolumn,showpacs,prb]{revtex4}%

\usepackage{amsfonts}
\usepackage{amsmath}
\usepackage{amssymb}
\usepackage{graphicx}

\begin{document}

\title{Anomalous Rashba spin-orbit interaction in InAs/GaSb quantum wells}
\author{Jun Li}
\author{Kai Chang}
\altaffiliation[Corresponding author:]{kchang@red.semi.ac.cn}
\affiliation{SKLSM, Institute of Semiconductors, Chinese Academy of Sciences, P. O. Box
912, Beijing 100083, China}
\author{G. Q. Hai}
\affiliation{Instituto de F\'{\i}sica de S\"{a}o Carlos, Universidade de S\"{a}o Paulo,
13560-970 S\"{a}o Carlos, S\"{a}o Paulo, Brazil}
\author{K. S. Chan}
\affiliation{Department of Physics and Materials Science, City University of Hong Kong,
Hong Kong, China}

\begin{abstract}
We investigate theoretically the Rashba spin-orbit interaction in InAs/GaSb
quantum wells (QWs). We find that the Rashba spin-splitting (RSS) depends
sensitively on the thickness of the InAs layer. The RSS exhibits nonlinear
behavior for narrow InAs/GaSb QWs and the oscillating feature for wide
InAs/GaSb QWs. The nonlinear and oscillating behaviors arise from the
weakened and enhanced interband coupling. The RSS also show asymmetric
features respect to the direction of the external electric field.
\end{abstract}

\maketitle

InAs/GaSb superlattices (SLs) and quantum wells (QWs) have attracted
intensive attention in the past decades due to their potential application
in nanoelectronics and remarkable electronic properties, e.g., the infrared
detector and laser diode, as well as interband tunneling diodes and
transistors\cite{Esaki,McGill,Ting1,Houng}. An interesting feature of this
broken-gap structure is that the top of the valence band of GaSb lies above
the bottom of the conduction band of InAs. A two-dimensional electron gas
(2DEG) in the InAs layer can coexist with a two-dimensional hole gas in the
GaSb layer since the electron can move across the InAs/GaSb interface from
the valence band of GaSb to the conduction band of InAs, consequently
leading to a semimetallic phase\cite{Esaki, JLuo}. The energy spectrum
exhibits an anticrossing behavior between the top valence subband and the
lowest conduction subband at finite in-plane momentum when the lowest
conduction subband in InAs layer lies below the top valence subband in GaSb
layer\cite{BandStructure}. The hybridized gap caused by the anticrossing was
observed experimentally\cite{GapExperiment}, and leads to the semiconducting
behavior of system. Recently, by utilizing the unique characteristics of the
InAs/GaSb/AlSb system, e.g., the strong spin-orbit interaction (SOI) in InAs
and GaSb and the high electron mobility of InAs, it may be possible to
realize high-speed spintronic devices, e.g., Rashba spin filters\cite%
{Vosk,Koga,Ting2}, a spin field effect transistor\cite{Datta} and a
high-frequency optical modulator utilizing spin precession\cite{Hallstein}.
The SOI also has a significant influence on the spin relaxation of electrons
in semiconductors and could be used to generate the spin current.

In this Letter, we investigate theoretically the Rashba spin-splitting (RSS)
in undoped InAs/GaSb quantum wells sandwiched by AlSb barriers by solving
the eight-band Kane Hamiltonian\cite{Burt,RSS} and Poisson equation
self-consistently. We find a spontaneous RSS that arises from the interface
contribution in the absence of external electric field, and RSS is
asymmetric with respect to the direction of external electric field. It is
interesting to note that RSS exhibits distinct behavior, i.e., nonlinear and
oscillating feature as a function of the in-plane momentum, at small and
large thicknesses of InAs layers due to the interband coupling and
hybridization between the conduction band and the valence band.

\begin{figure}[h]
\includegraphics[width=\columnwidth]{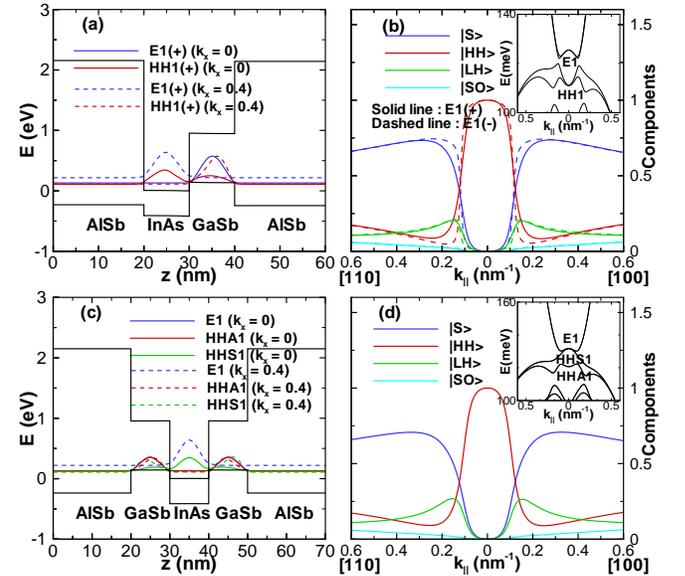}
\caption{(a) Band profile and the probability density distribution
of E1(+) (blue), HH1(+) (red) subband in InAs/GaSb ASQW at
$k_{\parallel }=0$ (solid line) and $k_{\parallel }=0.4~nm^{-1}$ (dashed line), ($L_{InAs}=10~nm$,$%
L_{GaSb}=10~nm$). (b) The four band-edge state components,
$|S\rangle $ (blue), $|HH\rangle $ (red), $|LH\rangle $ (green) and
$|SO\rangle $ (cyan) as a function of in-plane wavevector of E1
band. The solid and dashed line denote the spin up (E1(+)) and spin
down(E1(-)) state, respectively. The inset shows the calculated band
structure. (c) and (d), the same as (a) and (b), but for a
GaSb/InAs/GaSb SQW.} \label{fig:fig1}
\end{figure}

We consider an undoped InAs/GaSb quantum well (grown on the [001] plane)
shown schematically in Fig. 1(a). If an external electric field $F$ is
applied along the direction perpendicular to the QW plane, an external
electric field term $V_{E}\left( z\right) =eFz$ should be added to the total
Hamiltonian. When the layer thickness or the external electric field is
sufficiently large so that the bottom of the lowest conduction subband in
InAs layer falls below the top of the highest valence subband in GaSb layer,
electrons could transfer from GaSb layer to InAs layer, thus induced an
internal electrostatic potential $V_{in}\left( z\right) $ in InAs and GaSb
layers\cite{Bastard}. The total Hamiltonian writes as $H=H_{k}+V_{E}\left(
z\right) +V_{in}\left( z\right) $ including the external and internal
electrostatic potential. A self-consistent iteration procedure is performed
until $V_{in}\left( z\right) $ is stable. The Kane parameters of materials
used in our calculation are obtained from Ref. \onlinecite{Parameters}.
\begin{figure}[h]
\includegraphics[width=\columnwidth]{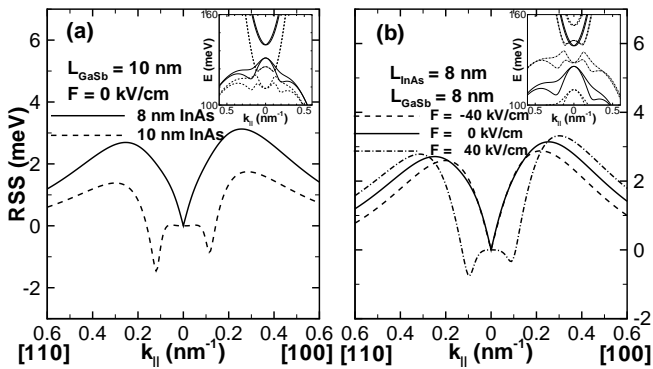}
\caption{(a)Rashba spin-splitting of E1 band as a function of the
in-plane momentum in the absence of external electric field for
different thicknesses of InAs layer. (b) The same as (a), but for
different external electric fields. The insets depict the band
structure near the band gap.} \label{fig:fig2}
\end{figure}

Fig. \ref{fig:fig1} (a) shows schematically the self-consistent band
profile of a 10 nm-10 nm InAs/GaSb asymmetric chemical multilayer QW (ASQW) at $%
k_{\parallel }=0$. The energy dispersion of the ASQW obtained from the
self-consistent calculation\ is plotted in the inset of Fig. \ref{fig:fig1}%
(b). From the energy dispersion, we can find that the energy dispersion of
the lowest (highest) conduction (valence) subband shows a minimum (maximum)
at a finite $k_{\parallel }^{a}$, i.e., anticrossing behavior, since the
bottom of the lowest electron subband in the InAs layer lies below the top
of the highest heavy-hole subband in the GaSb layer. A spin-dependent
hybridized gap($\sim $3 meV) at the anticrossing point forms due to the
strong mixing effect of the InAs electron state and the GaSb hole state. In
this QW, the concentration of electrons transferred from GaSb layer to InAs
layer is found to be at the order of 1.9 $\times $ 10$^{-11}$ cm$^{-2}$,
which induces a 7 meV bending down (up) in the profile of InAs's conduction
band (GaSb's valence band) near the interface. The small band-bending only
change the results slightly and shifts $k_{\parallel }^{a}$ to a smaller
value. In Fig. \ref{fig:fig1}(b), we plot the four components of the E1
state as a function of $k_{\parallel }$. The dominant component of E1 states
(E1(+) and E1(-)) experiences a crossover from $|HH\rangle $ to $|S\rangle $
when the in-plane momentum $k_{\parallel }$ sweeps across the anticrossing
point $k_{\parallel }^{a}$. Meanwhile, the dominant component of the HH1
band varies from electron-like to hole-like feature. This feature can also
be demonstrated in Fig. \ref{fig:fig1}(a) in which we also plot the density
distribution of the E1(+) and HH1(+) at $k_{\parallel }=0$ and $0.4~nm^{-1}$%
. At $k_{\parallel }=0$, the E1 (HH1) state is mostly heavy-hole-like
(electron-like) and therefore localizes in the GaSb (InAs) layer. At $%
k_{\parallel }>k_{\parallel }^{a}$, the two anticrossing subbands E1 and HH1
exchange their main characteristics, so the density distribution of E1(HH1)
state localizes in InAs(GaSb) layer. Fig. \ref{fig:fig1} (c) and (d) plot
the situation of a symmetric chemical multilayer QW (SQW). In this symmetric
structure, the internal electrostatic potential is symmetric respect to the
center of InAs layer. Thus there is no RSS existing, and the components of
the two spin branches of E1 are identical in Fig. 1(d).

In Fig. \ref{fig:fig2} (a) we plot the Rashba spin-splitting (RSS)
of the lowest conduction subband (E1) as a function of the in-plane
momentum at fixed thicknesses of the InAs and GaSb layers. As shown
in the insets, the energy bands of narrow QW, e.g., 8 nm InAs layer
with 10 nm GaSb layer, manifest a normal semiconductor phase since
the strong quantum confinement pushes the lowest conduction subband
to a higher energy. The RSS in this QW is a nonlinear function of
in-plane wave vector, just as the RSS in biased narrow band gap
semiconductor QWs we reported elsewhere before \cite{RSS}.
Interesting difference between this work and the previous work
\cite{RSS} is that there exist a spontaneous RSS in the InAs/GaSb QW
in the absence of external electric fields, since it arises mainly
from the asymmetric potential profile of InAs/GaSb QW, i.e., the
asymmetric interband coupling at the left and right interfaces for
InAs and GaSb layers. This is clearly demonstrated by the
disappearing of the spontaneous RSS in the GaSb/InAs/GaSb SQW (see
Fig. \ref{fig:fig1}(c) and (d)). Besides the nonlinear RSS, one can
also see an oscillating RSS of the E1 band in a wider QW, e.g., the
QW of 10 nm InAs layer with 10 nm GaSb layer, in which an
anticrossing occurs in the energy spectrum (see the inset of Fig. \ref%
{fig:fig2} (a)). A sharp drop of RSS corresponding to the anticrossing point
$k_{\parallel }^{a}$ appears. It is interesting to see that the RSS changed
its sign near the anticrossing point. Obviously the sign change of RSS
corresponds to the cross of the two spin branches. In addition, the RSS is
anisotropic with respect to different $k_{\parallel }$ direction, e.g.,
[100] and [110] direction (see Fig. 2). The anisotropy of RSS comes from the
anisotropy of valence bands of InAs/GaSb QW through interband coupling.
\begin{figure}[h]
\includegraphics[width=\columnwidth]{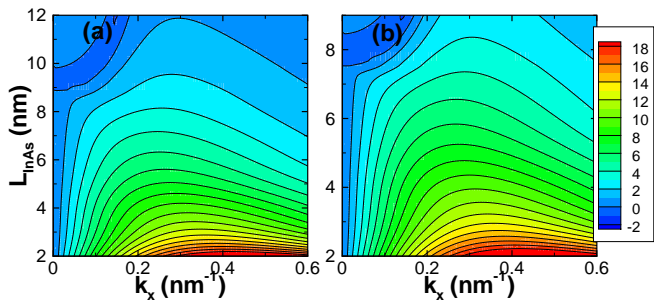}
\caption{(a) Contour plot of self-consistently calculated RSS of E1
band as a function of the in-plane momentum and the thickness of the
InAs layer without external electric field. (b) The same as (a), but
with an external electric field $F=40~kV/cm$. In each panel the
thickness of GaSb layer is fixed at 8 nm.} \label{fig:fig3}
\end{figure}

Fig. \ref{fig:fig2} (b) displays the RSS of an InAs/GaSb quantum
well as a function of the in-plane momentum for different external
perpendicular electric fields. This figure shows that the RSS is
heavily controlled by the external electric field. This behavior of
RSS can be explained by the interplay between the asymmetric
interface contribution and the interband coupling induced by the
external electric field. In Fig. \ref{fig:fig2} (b), if an external
electric field is applied parallel to the positive direction of the
$z$ axis, the energy of the conduction subbands decreases while the
energy of the valence subbands increases. Thus, the anticrossing
behavior occurs or is enhanced even for QW with a narrow width InAs
layer that
exhibits a normal-semiconductor phase in zero electric field\cite%
{EfieldEffect}. A valley of RSS appears once an anticrossing happens. The
energy difference between the E1 and HH1 subbands is increased when the
external electric field is applied antiparallel to the z axis. Therefore the
anticrossing behavior is weakened and even smeared out. Compared to the
conventional type-I QW, RSS in the type-II InAs/GaSb broken-gap QW exhibits
unique features, i.e., nonlinear and oscillating behaviors which can be
tuned by the external electric field.

For the wide ASQW case (see Figs. \ref{fig:fig2} (a) and (b)), the internal
electrostatic potential induced by charge transfer tends to weaken the
coupling between the conduction subbands in InAs layer and the valence
subbands in GaSb layer, i.e., the internal electric field compensates partly
the external electric field, and shifts the anticrossing point to a smaller $%
k_{\parallel }$. In normal semiconductor phases of undoped InAs/GaSb QW, the
charge transfer process doesn't happen, therefore the internal electrostatic
potential disappears.

Figs. \ref{fig:fig3} (a) and (b) describe the RSS as function of the
in-plane momentum and the thickness of the InAs layer in zero and finite
electric field. From this contour plot one can see more clearly that the RSS
shows a nonlinear feature for the narrow InAs layer, and an oscillating
behavior at large thickness of InAs layer. The nonlinear behavior arises
from the interface contribution which also depends on the interband coupling%
\cite{RSS}. The oscillation of RSS is caused by the strong mixing between
the conduction subband E1 and the heavy-hole subband HH1. Interestingly,
this oscillating behavior of RSS\ can be enhanced by an electric field,
e.g., $F=40~kV/cm$, (see Fig. \ref{fig:fig3} (b)). Note that there is a
critical thickness of the InAs layer $L_{c}$, the RSS exhibits oscillating
features corresponding to the different phases when $L>L_{c}$.

\begin{figure}[h]
\includegraphics[width=\columnwidth]{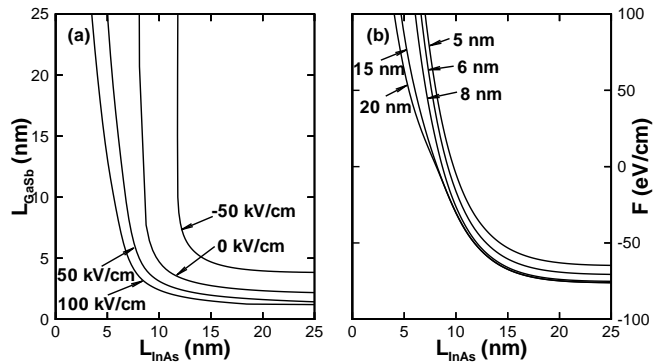}
\caption{(a) The phase diagram of InAs/GaSb QWs as function of the
 thickness of InAs and GaSb layers for different electric fields. (b)
The Same as (a), but as function of the thickness of InAs layer and
the external electric field for different thicknesses of the GaSb
layers.} \label{fig:fig4}
\end{figure}

Fig.\ref{fig:fig4} (a) displays the phase diagram of InAs/GaSb QWs for
different external electric fields. This figure indicates that the critical
thickness of the InAs layer decreases as the thickness of GaSb layer
increases, and tends to saturate at different thicknesses determined by the
external electric fields (see Fig. \ref{fig:fig4} (a)). Positive external
electric fields decrease the critical thickness, while negative external
electric fields increase the critical thickness. Fig. \ref{fig:fig4} (b)
gives the critical thickness of InAs layer and the critical electric field
for different thicknesses of the GaSb layers. The critical electric fields
saturate at different valyes for different thicknesses of the GaSb layers.

In summary, we theoretically investigated Rashba spin-orbit
interaction in InAs/GaSb asymmetric chemical multilayer QWs. We
found a spontaneous RSS that arises from the interface contribution
induced by the asymmetric structure of the QW. The RSS exhibits
distinct behaviors, i.e., nonlinear and oscillating behavior,
depending on the thickness of the QW and the external electric
field. The oscillating RSS comes from the strong interband mixing
between the lowest InAs conduction and the highest valence GaSb
subbands. This crossover between two distinct behaviors can be tuned
by the thicknesses of the InAs or GaSb layers and the external
electric field. The unique features of RSS in InAs/GaSb QWs could
provide us an interesting way to manipulate the electron spin and
construct spintronic devices.

\begin{acknowledgments}
This work was supported by the NSFC Grant No. 60525405 and the
knowledge innovation project from CAS, , City University of Hong
Kong Strategic Research Grant (project no. 7002029). GQH was
supported by FAPESP and CNPq (Brazil).
\end{acknowledgments}

\end{document}